\documentclass{aastex}
\usepackage{emulateapj5,times,mathptm}

\newcommand{\simgt}{\lower 2pt \hbox{$\, \buildrel {\scriptstyle >}\over {\scriptstyle\sim}\,$}}
\newcommand{\simlt}{\lower 2pt \hbox{$\, \buildrel {\scriptstyle <}\over {\scriptstyle\sim}\,$}}

\newcommand{\apm}{APM~08279+5255}

\newcommand{\asca}{{\emph{ASCA}}}
\newcommand{\chandra}{{\emph{Chandra}}}

\newcommand{\xmm}{\emph{XMM-Newton}}

\newcommand{\hst}{{\emph{HST}}}

\slugcomment{Received 2002 May 30; accepted 2002 July 8}
\shorttitle{ Relativistic X-ray BALs from APM~08279+5255}
\shortauthors{CHARTAS ET AL.}

\tighten

\begin{document}

\def\sarc{$^{\prime\prime}\!\!.$}
\def\arcsec{$^{\prime\prime}$}
\def\beginrefer{\section*{References}%
\begin{quotation}\mbox{}\par}
\def\refer#1\par{{\setlength{\parindent}{-\leftmargin}\indent#1\par}}
\def\endrefer{\end{quotation}}

\title{ {\sl CHANDRA} Detects Relativistic Broad Absorption Lines from \apm}

\author{G. Chartas,\altaffilmark{1} W. N. Brandt,\altaffilmark{1} S. C. Gallagher,\altaffilmark{1,2}
and G. P. Garmire\altaffilmark{1}}

\altaffiltext{1}{Astronomy and Astrophysics Department, Pennsylvania State University,
University Park, PA 16802. chartas@astro.psu.edu, niel@astro.psu.edu, gallsc@astro.psu.edu,
garmire@astro.psu.edu}
%\altaffiltext{2}{MIT Center for Space Research, 70 Vassar Street, Cambridge, MA 02139.}

\altaffiltext{2}{MIT Center for Space Research, 77 Massachusetts Avenue, Cambridge, MA 02139.}

\begin{abstract}

We report the discovery of X-ray broad absorption lines (BALs)
from the BALQSO \apm\ originating from material moving at relativistic velocities
with respect to the central source. The large flux magnification by a factor of $\sim$ 100
provided by the gravitational lens effect combined with the
large redshift ($z = 3.91$) of the quasar have 
facilitated the acquisition of the first high signal-to-noise X-ray spectrum 
of a quasar containing X-ray BALs.
Our analysis of the X-ray spectrum of \apm\ 
places the rest-frame energies of the two observed absorption lines at 8.1 and 9.8~keV. 
The detection of each of these lines is 
significant at the $ > $ 99.9\% confidence level based on the $F$-test.
Assuming that the absorption lines are from
%ion responsible for the absorption 
Fe~{\sc xxv} K${\alpha}$, the implied bulk velocities of the
X-ray BALs are $\sim$ $0.2c$ and $\sim$ $0.4c$, respectively.
%With the present observation we address several fundamental questions 
%related to the  BAL phenomenon;
The observed high bulk velocities of the X-ray BALs 
combined with % the high degree of ionization and 
the relatively short recombination time-scales of the X-ray absorbing gas
imply that the  absorbers responsible for the X-ray 
BALs  are located at radii of \simlt 2 $\times$ 10$^{17}$~cm,
within the expected location of the UV absorber. 
With this implied geometry the X-ray gas could provide the necessary shielding 
to prevent the UV absorber from being completely ionized by the central X-ray source, 
consistent with hydrodynamical simulations of line-driven
disk winds. Estimated mass-outflow rates for the gas creating the X-ray BALs are typically
less than a solar mass per year. 
%if the X-ray BALs are located at distances of 
%\simlt 2 $\times$ 10$^{17}$~cm, consistent with our previously inferred 
%radii of the X-ray BALs. 
%Constraining this value is essential for understanding
%the connection between black hole and bulge growth in the host
%galaxy. 
Our spectral analysis also indicates 
that the continuum X-ray emission of \apm\ is consistent with that 
of a typical radio-quiet quasar with a spectral slope of $\Gamma$~=~1.72$_{-0.05}^{+0.06}$. 
%and an underlying column density of about 7 $\times$ 10$^{23}$~cm$^{-2}$.

\end{abstract}

\keywords{galaxies: active --- quasars: absorption lines --- quasars: individual~(APM~08279+5255) --- X-rays: galaxies --- gravitational lensing} 

\section{INTRODUCTION}

It is commonly accepted that most quasars contain energetic outflows of ionized gas
emerging from their accretion disks at speeds
ranging from $\approx$ 5,000--30,000~km s$^{-1}$ (e.g., Turnshek et al. 1988; Weymann et al. 1991).  
These outflows imprint broad absorption features bluewards 
of the resonant UV emission lines of C~{\sc iv}, Si~{\sc iv}, N~{\sc v}, and O~{\sc vi}. 
Broad absorption features are expected
to be observed only for lines of sight that traverse the outflow.
The outflow is thought to be driven by radiation 
pressure on spectral lines from UV photons of the central source 
(e.g., Arav et al. 1995; Murray et al. 1995; Proga, Stone, \& Kallman 2000; 
Srianand et al. 2002).
An estimate of the mass-outflow rate may result from the study 
of the properties of the outflowing winds of quasars. 
This quantity may be used to evaluate the contribution of
outflowing winds in distributing accretion-disk material into 
the vicinity of the quasar central engine and into the host galaxy
over a typical life time of a quasar. Constraining this rate is 
also important for understanding the connection 
between black hole and bulge growth in the host galaxy (e.g., Fabian 1999).

An estimate of the mass-outflow rate requires knowledge of 
the velocities and locations of the various ions that contribute
to the wind. 
Broad absorption features in the UV band often show
multiple detached components with different velocities,
column densities, and ionization states. One needs to include 
the contributions of all components to obtain an accurate 
value of the total mass-outflow rate.
Present estimates of mass-outflow rates are based mostly on 
the contributions from ions absorbing in the rest-frame UV band.
The present X-ray data for BALQSOs are sparse, and only poor 
to moderate signal-to-noise ratio (S/N) spectra are 
available (e.g., Chartas et al. 2001; Gallagher et al. 2001; 
Green et al. 2001; Oshima et al. 2001; Brinkmann et al. 2002; 
Gallagher et al. 2002). The few moderate S/N X-ray spectra of BALQSOs available 
show that their X-ray faintness is due to
absorption with typical hydrogen column densities ranging from
$\sim$ 10$^{22}$--10$^{24}$~cm$^{-2}$. The ionization, kinematic and 
spatial properties of the X-ray absorbing material
are not well constrained. The physical relationship between the 
UV and X-ray absorbers is unclear. Is the X-ray absorber part of the outflow?
%Another unresolved issue is the presence of UV spectral lines 
%of moderate-to-high ionization in BALQSOs 
%in the presence of the observed extreme UV and soft X-ray photons
%produced by the central source.
Another unresolved issue is how moderately ionized species
can survive the extreme UV and soft X-rays
produced by the central source.
To account for this, theoretical studies have postulated
the presence of an optically thick layer of shielding gas between
the central source and the outflow that prevents the outflow from
becoming completely ionized (e.g., Murray et al. 1995). 
Recent simulations by Proga et al. (2000) indicate that the outflow is self-shielding,
i.e., the shielding gas is an integral component of the outflow.
The observed X-ray absorbing gas in BALQSOs has been
suggested as a candidate for the shielding gas;
however, there has been little direct observational evidence
for this association. Recently, narrow absorption lines (NALs) in the X-ray band
have been detected in \chandra\ and \xmm\ observations of bright
Seyfert 1 galaxies (e.g., Kaspi et al. 2002). The 
NALs are blueshifted relative to the systemic velocity suggesting that the NAL material
is part of an outflow with a mean outflow velocity of a few hundred km~s$^{-1}$. 

To improve our understanding of the X-ray absorption in BALQSOs, we
performed a long \chandra\ observation of the bright gravitationally lensed
BAL quasar \apm. The flux magnification of \apm, estimated to be $\sim$
100 (Egami et al. 2000; Mu{\~ n}oz et al. 2001) in the X-ray band, and 
its high redshift of $z = 3.91$ allowed us to study the kinematic and 
ionization properties of a BAL quasar in the X-ray band. Specifically, the
high redshift of \apm\ places the strong Fe~K features at energies where the
telescope effective area is much larger. Here we present the results of these observations.

\section{OBSERVATION AND DATA REDUCTION}

\apm\ was observed as part of the guaranteed time observer program with
the Advanced CCD Imaging Spectrometer
(ACIS) instrument (G. P. Garmire et al. 2002, in preparation) onboard the {\it Chandra X-ray
Observatory} (CXO) on 2000 October 11 and on 2002 February 24
for 9.2~ks and 88.8~ks, respectively. The pointing of the telescope placed 
\apm\ on the back-illuminated S3 chip of ACIS.
In Table 1 we list the observation epochs, exposure times and
detected event rates for the lensed images (see~$\S$~3).
For preparing the \chandra\ data for analysis we used the CIAO 2.2 and CALDB 2.12 products
provided by the \chandra\ X-ray Center (CXC).
To improve the spatial resolution we
removed a $\pm$ 0\sarc25 randomization applied to the event positions
in the CXC processing and employed a sub-pixel resolution technique
developed by Tsunemi et al. (2001).
To account for the recently observed quantum efficiency decay of ACIS, possibly caused by
molecular contamination of the ACIS filters, we have applied a time-dependent correction to the 
ACIS quantum efficiency based on the presently available information
from the CXC. The ACIS quantum efficiency decay is insignificant for energies above 1~keV 
and does not affect the main results of our analysis.

\normalsize

\section{RELATIVE ASTROMETRY AND PHOTOMETRY}
The \chandra\ image of \apm\ obtained from the 88.8~ks observation is
shown in Figure 1. To improve the spatial resolution we have 
applied the Lucy-Richardson (L-R) maximum likelihood
deconvolution technique (Richardson 1972; Lucy 1974). For the deconvolution we supplied a
point spread function (PSF) created by the simulation tool \verb+MARX+ (Wise
et al. 1997).  The X-ray spectrum used to generate the PSF was that
determined from our spectral analysis (see $\S$ 4).
We find a separation between the X-ray images A and B of 0\sarc38 $\pm$ 0\sarc01.
Recent observations of \apm\ with the NICMOS camera onboard 
\hst\ (Ibata et al. 1999) imply the presence of three
images with a separation between the two brightest 
images A and B of $\Delta\theta_{AB}$=0\sarc38   
and between A and the fainter image C of $\Delta\theta_{AC}$=0\sarc1.
To estimate the X-ray flux ratios we modeled the \chandra\ images of A, B and C with 
point spread functions generated by the simulation tool \verb+MARX+.
The X-ray event locations were binned with a bin-size of 0\sarc0246.
The simulated PSFs were fit to the \chandra\ data by minimizing the
Cash $C$ statistic formed between the observed and simulated images
of \apm. The relative positions of the images were fixed to
the observed NICMOS values.  We find X-ray flux ratios of
$[f_{C}/f_{A}]_{X} = 0.21 \pm 0.02$ and $[f_{B}/f_{A}]_{X} = 0.88 \pm 0.03$. 
The X-ray flux ratios are close to the flux ratios measured in the 
\hst\ F160 band of $ [f_{C}/f_{A}]_{F160} = 0.22 \pm 0.01 $ 
and $ [f_{B}/f_{A}]_{F160} = 0.78 \pm 0.01$. 
We briefly comment that X-ray flux variability is detected in both images of \apm,
and this issue will be the focus of a future paper.

\section{SPECTRAL ANALYSIS}
Spectra of \apm\ from the combined images for the two epochs
were first extracted and analyzed separately to search for any 
significant long-term variability.  
The source spectra were extracted from circular regions
centered on the midpoint between images A and B with radii of 3\arcsec.
The background was determined by extracting events within a source-free
annulus centered on the midpoint between images A and B 
with inner and outer radii of 10\arcsec\ and
30\arcsec, respectively. All derived errors below are at the 90\%
confidence level unless quoted otherwise. 
Results from spectral fits to the 9.1~ks observation
of \apm\ were reported in Gallagher et al. (2002).
The X-ray flux of \apm\ appears to have decreased by a factor
of $\sim$ 2 between the first and second epochs (see Table 1).
Unfortunately, there was a moderate degree of photon pile-up in the first observation of 
\apm\ which can lead to spectral distortions and loss of events. 
We therefore did not combine the spectra of the first and second epochs. 
The second observation of \apm\ was performed
in a sub-array mode to reduce the effects of pile-up. 
The reduced X-ray flux of \apm\ during the second epoch and the reduced
CCD frame-time of 2.5 s used for this observation significantly reduced
the effects of pile-up, which is
negligible in the subsequent analysis. In the spectral analysis that follows
we focus on the data obtained from the 88.8~ks 
observation of \apm.  

A variety of models were fit to the 88.8~ks spectrum of \apm\
employing the software tool XSPEC v11 (Arnaud 1996).
The spectrum was initially fit with a  
model consisting of a simple power law with
Galactic absorption due to neutral cold gas with a column density
of $N_{\rm H}$ = 3.9 $\times$ 10$^{20}$~cm$^{-2}$ (Stark et al. 1992).
The model also included neutral intrinsic absorption
at $z = 3.91$. Our fits support the presence of an intrinsic absorber
with a column density of $N_{\rm H}$ = (6.0$_{-0.8}^{+0.8}$) $\times$ 10$^{22}$~cm$^{-2}$ 
(see Table 2).
The fit is not acceptable in a statistical sense with 
$\chi^{2}$ = 182.1 for 109 degrees of freedom (dof).
%To obtain insight into the nature of the
%intrinsic absorber the data below 10.8~keV in the
%rest frame were ignored, and the spectrum was fit with only a
%power-law model modified by Galactic absorption. 
%The presence of intrinsic absorption was initially
%revealed by fitting the \apm\ data above
%the rest frame energy of 10.8~keV with a
%simple power-law model modified by Galactic absorption
%and extrapolating this model to lower energies.
The fit residuals show several absorption features between 1.5--3~keV 
that contribute to the unacceptable fit.
To illustrate better the presence of these absorption features
and the absorption at energies below 1~keV, we fit the \apm\ data 
above the rest frame energy of 
10.8~keV with a simple power-law model
modified by Galactic absorption
and extrapolated this model to lower energies.
The spectrum above 10.8~keV (rest-frame) is less susceptible
to intrinsic absorption and excludes the
absorption features.
Following this analysis strategy, the low-energy
residuals indicate that intrinsic absorption is present in
the spectrum. In addition, the spectrum shows two strong absorption
lines from 1.5--3~keV; these are presented in Figure 2a.
We attempt to model the residual features by considering a range of models of 
increasing complexity. 
We first add to our absorbed power-law model a redshifted Gaussian component
near the most obvious absorption feature appearing at an energy of $\sim$ 
1.65~keV (observed frame). Including one Gaussian component in our model leads to a significant 
improvement in fit quality at the $ > $~99.9\% confidence level (according to the $F$-test)
with $\chi^{2}$ = 146.9 for 106 dof. The best-fit energy of the absorption feature 
is $E_{abs1}$ = 8.05$_{-0.07}^{+0.18}$~keV (rest frame).
This absorption feature is in a well-calibrated
energy region, and the combined effective area of the \chandra\
mirrors and ACIS is known to vary smoothly near this energy.
The absorption feature is not resolved by ACIS; we place an upper limit 
on its width of 140~eV (corresponding to a FWHM of $\simlt$ 12,300 km s$^{-1}$)
at the 90\% confidence level.
We next add a second Gaussian component near 1.9~keV to model the remaining residuals
near this energy. This fit is significantly improved compared to the previous fit 
at the $ > $ 99.9\% confidence level
and yields $\chi^{2}$ = 106.7 for 103 dof; the fit is now statistically acceptable. 
The best-fit value for the photon index is $\Gamma = 1.72_{-0.05}^{+0.06}$,
consistent with the range of $\Gamma$ measured for large samples of radio-quiet quasars 
at lower redshifts (e.g., George et al. 2000; Reeves \& Turner 2000).  
The best-fit energy for the second absorption line is $E_{abs2}$ = 9.79$_{-0.19}^{+0.20}$~keV with a width
of $\sigma_{abs2}$ = 0.41$_{-0.16}^{+0.19}$~keV. The second absorption feature 
falls near the instrumental iridium edge produced by the \chandra\ mirrors. 
To ascertain any systematic calibration uncertainties near the mirror edge
we have fit the spectra of several  
test-sources\footnote{The test-sources observed with 
ACIS S3 are the supernova remnant G21.5-0.9
observed on 2001 March 18, the millisecond pulsar J0437-4715
observed on 2000 May 29 and the radio-loud quasar Q0957+561 observed
on 2000 April 16.} with expected smooth power-law spectra 
near the 2~keV iridium edge. 
Since $\chi^{2}$ residuals depend on the statistics, we filtered the test-source
data in time to produce spectra with a total
number of counts equal to that observed in the second
observation of \apm. Typical $\chi^{2}$ residuals for these 
test-sources near the mirror edge 
are less than $\sim$ 1$\sigma$ indicating that the observed 
5$\sigma$ residuals near 2~keV are real and not
due to systematic errors in the calibration of the effective area of
the \chandra\ mirrors. We find a ratio of $\sim$ 0.5 between data and model near
the two absorption features for the second observation of \apm.
This ratio is significantly less than
the observed ratio of $\sim$ 0.95 between data and model
for our test-sources.
The fit that includes two Gaussian absorption-line components
shows positive residuals near 8.4~keV and 11.2~keV (rest frame). 
These residuals may at least partially be the result of the simplistic absorption-line models
that we have adopted in the present analysis. UV BALs usually show  
multiple absorption components with non-Gaussian profiles, and such complex profiles  
may also be present in X-ray BALs. 
To test further the robustness of our modeling
we also attempted to fit the 1.5--3~keV residuals in Figure 2a
with a model that included an absorption edge. Specifically, for a model 
consisting of a simple power law with Galactic absorption, intrinsic
absorption, and one absorption edge we find a best fit energy 
and optical depth of the edge of $E_{edge1}$ = 7.68$_{-0.25}^{+0.21}$~keV
and $\tau_{edge1}$ = 0.37$_{-0.13}^{+0.14}$, respectively.
This model did not provide an acceptable fit
with ${\chi}^{2}$ = 146 for 107 dof.
In Figure 2b we show that the fit of the \apm\ spectrum with a one-edge model
produces significant residuals. 
The reason for these large residuals is that 
the absorption edge is too broad to fit the narrow absorption lines.
The addition of a second absorption edge
to the model did not result in a significant improvement of the fit.

\section{DISCUSSION AND CONCLUSIONS}

Our analysis of the \chandra\ spectrum of \apm\ shows strong evidence for
the presence of absorption lines at rest-frame energies of 8.05$_{-0.08}^{+0.10}$~keV and 
9.79$_{-0.19}^{+0.20}$~keV (fit~3 of Table 2). 
Of all the abundant elements, iron absorption lines would be
the closest in energy to the observed features. In this sense, 
our interpretation that the absorption lines are associated
with Fe~K absorption is the most conservative one possible
(e.g., absorption lines from relativistic oxygen would require 
much larger blueshifts). At higher atomic numbers than iron
(corresponding to higher absorption-line energies), 
there are no abundant elements. Observationally,
Fe~K absorption lines have been seen from other objects,
such as the X-ray binaries GRS~1915+105 (Kotani et al. 2000), 
Circinus~X-1 (Brandt \& Schulz 2000), and GX~13+1 (Ueda et al. 2001)
and possibly the AGN NGC~3516 (Nandra et al. 1999). 
We note that the expected ratios of the energies of the 
iron K$\beta$ to K$\alpha$ transitions of Fe~{\sc xxv} and Fe~{\sc xxvi}
are just outside the 2$\sigma$ confidence limits of the 
ratio of the energies of the two observed absorption lines.
Because of the possible presence of multiple components within each of the 
observed absorption lines, we expect that future X-ray observations of \apm\ with 
higher energy resolution will unambiguously show whether
the observed absorption lines correspond to the K$\alpha$ and K$\beta$ transitions of Fe~{\sc xxv}
or Fe~{\sc xxvi}.

A plausible site that may be producing the absorption features
is the outflowing disk wind. Complex absorption features
are also observed in the UV spectrum of \apm. In particular,
recent high resolution spectroscopy of \apm\ reveals
multiple velocity components of C~{\sc iv} $\lambda$(1548~\AA, 1551~\AA), 
at $\sim$ 4670~km~s$^{-1}$, $\sim$ 9670~km~s$^{-1}$, and $\sim$
12,400~km~s$^{-1}$ (Srianand \& Petitjean 2000). 
Broad absorption due to O~{\sc vi} $\lambda$(1032~\AA, 1037~\AA), 
N~{\sc v}$\lambda$(1238~\AA, 1242~\AA) and 
Si~{\sc iv} $\lambda$(1393~\AA, 1402~\AA) is also detected.  
The wide range of ionization levels, velocities and 
column densities inferred from the UV BALs
of \apm\ imply that the wind may be composed of multiple regions
of different densities, with large ionization gradients (Srianand \& Petitjean 2000).
For our estimate of the ejection velocities of the absorbers
we assume as argued above that the X-ray absorption features are due to 
absorption from Fe ions in the flow.
The rest energies of the most likely resonant absorption lines of Fe
are 6.70~keV (Fe~{\sc xxv} K$\alpha$), 7.88~keV (Fe~{\sc xxv} K$\beta$),
6.97~keV (Fe~{\sc xxvi} K$\alpha$) and 8.25~keV (Fe~{\sc xxvi} K$\beta$).
We estimate that the 8.05~keV and 9.79~keV
absorption features correspond to wind velocities (depending on the ionization state of Fe)
of $0.20c$~(Fe~{\sc xxv}~K$\alpha$), 
$0.15c$~(Fe~{\sc xxvi}~K$\alpha$) and $0.40c$~(Fe~{\sc xxv}~K$\alpha$), 
$0.36c$~(Fe~{\sc xxvi}~K$\alpha$) 
respectively, relative to the systemic redshift of $z = 3.91$.
For the velocity calculations we considered the special relativistic 
velocity correction and assumed that the angle between the wind velocity and 
our line of sight is 20$^{\circ}$.  This angle is not constrained with
the present data; however, hydrodynamical simulations indicate that the BAL wind
opening angle may range between 10$^{\circ}$--30$^{\circ}$
depending on the location of the inner radius of the disk.
We examined the low-resolution optical spectrum of \apm\
presented in Ellison et al. (1999) for evidence of notable absorption in
C~{\sc iv}, Si~{\sc iv}, and N~{\sc v} with velocities consistent with the
X-ray absorption lines.  In the spectral region covered, no strong systems
were apparent, though a detailed analysis of a high-resolution spectrum
(such as that presented in Srianand \& Petitjean 2000 for lower
velocities) remains to be done. The non-detection of these ions
is not a serious problem as they probably are not present in 
such a highly ionized gas.

To obtain some insight into the kinematics of the wind flow and the 
relative locations of the UV and X-ray absorbers,
we calculated the wind velocity as a function of radius.
More realistic velocity profiles of disk outflows have been 
obtained from hydrodynamical calculations (e.g., Proga et al. 2000)
that predict radial and azimuthal variations of the wind velocity.
For the purposes of our simple analysis 
we assume that the velocity of an outflow produced by radiation pressure 
from a central source with a UV luminosity of $L_{UV}$
and a mass of $M_{bh}$ is given by:
\begin{equation}
{ v_{wind} = \left [ 2GM_{bh}\left( {\Gamma_{f}}{{L_{UV}}\over{L_{Edd}}} - 1\right) \left( {{1}\over{R_{in}}} - {{1}\over{R}} \right) \right]^{1/2}},
\end{equation}

\noindent
where $L_{Edd}$ is the Eddington luminosity, $\Gamma_{f}$ is the force multiplier 
(see the discussion of $\Gamma_{f}$ in Laor \& Brandt 2002), 
$R_{in}$ is the radius at which the wind is launched from the disk, and $R$ is the distance of the
accelerated portion of the flow from the central source. 
In Figure 3 we plot wind velocity versus radius for material
launched at radii of 2 $\times$ 10$^{17}$~cm,
5 $\times$ 10$^{17}$~cm and 1 $\times$ 10$^{18}$~cm.
We have assumed ${\Gamma}_{f}$ = 100 (e.g., Arav, Li, \& Begelman 1994), 
$L_{UV}$ = 4 $\times$ 10$^{46}$ erg s$^{-1}$,
$L_{Bol}$ = 2 $\times$ 10$^{47}$ erg s$^{-1}$, and $L_{Bol}$/$L_{Edd}$ = 0.1.
Simulations performed by Proga et al. (2000) for a disk
accreting onto a 10$^{8}$ $M_{\odot}$ black hole yield
a minimum launching radius for the UV absorber of $\sim$ 10$^{16}$~cm.
Scaling this result for the likely black hole mass of \apm\ of $\sim$ 2~$\times$~10$^{10}$~$M_{\odot}$,
we estimate a minimum launching radius for the UV absorber of $\sim$ 2~$\times$~10$^{17}$~cm
for this luminous quasar.
From a qualitative perspective our velocity curves indicate that the 
wind reaches nearly its terminal velocity close to the launching radius.
A possible explanation of the relatively large velocities of the X-ray BALs 
compared to the UV BALs is that the 
material creating the X-ray BALs is launched at smaller radii;
the relatively small launching radii implied
are consistent with the higher degree of ionization needed to
produce the observed absorption lines if they are due to Fe~{\sc xxv} and/or Fe~{\sc xxvi}. 
A rigorous calculation of the velocity of the X-ray BAL wind is
beyond the scope of this letter. We note, however, that the 
largest unknown in this estimation is the value of ${\Gamma}_f$, 
which will depend on the ionization state and velocity
structure of the gas.  If the observed X-ray absorption is due to Fe {\sc xxv}, 
the high ionization parameter of this gas indicates that only
heavier metals will still have any electrons.  In this case, the dominant
radiation pressure may be provided by X-rays rather than UV photons.
Specifically, such highly ionized gas will be driven primarily by
bound-free absorption and Compton scattering, though considerable non-thermal broadening may
cause bound-bound absorption to contribute significantly as well
(e.g., Chelouche \& Netzer 2001; D. Chelouche 2002, private communication). 
A comparison between the recombination time-scale
of the ionized wind, $t_{recomb}$, and the travel-time, 
$t_{travel} = {\int_{R_{in}}^{R_{out}}{v_{wind}^{-1}dR}}$,
for the gas to reach a certain distance from the launching radius
can constrain the location of the material making the X-ray BALs.
For $t_{recomb}$ $ \ll $ $t_{travel}$
we expect the X-ray BAL material to be at small radii (near the launching radius)
to account for the high ionization. If $t_{recomb}$ $ \gg $ $t_{travel}$
the X-ray BAL can be located significantly away from the launching radius. 
The recombination time-scale for Fe~{\sc xxv} is 
$t_{recomb}$ $\sim 3 \times 10^{4} Z^{-2} T^{1/2}_{5} n^{-1}_{9}$~s, 
where $T = 10^{5}T_{5}$~K is the electron temperature 
and $n = 10^{9}n_{9}$~cm$^{-3}$ is the electron number density (Allen 1973).  
For a range of electron densities of 1 $\times$ 10$^{7}$~cm$^{-3}$
to 1~$\times$~10$^{10}$~cm$^{-3}$ (hydrodynamical models of line-driven disk winds
by Proga et al. 2000 predict electron number densities at the base of the wind of
5~$\times$~10$^{7}$~cm$^{-3}$ to 5~$\times$~10$^{9}$~cm$^{-3}$)
we find the recombination time-scales
to range between 4.4~$\times$~10$^{3}$~s and 4.4~s.
For a launching radius of 2 $\times$ 10$^{16}$~cm the amount
of time needed for the radiatively driven gas to reach a distance 
of 5 $\times$ 10$^{16}$~cm from the launching radius is $\sim$ 4 $\times$ 10$^{6}$~s.
We conclude that $t_{recomb}$~$ \ll $~$t_{travel}$
implying that the X-ray BAL material is located at relatively small
radii ( $ < $ 2 $\times$ 10$^{17}$~cm ) near the launching radius of the wind. Again this 
geometry explains both the 
high velocity of the observed X-ray BALs and the high ionization needed
to obtain Fe~{\sc xxv} or Fe~{\sc xxvi}.

Using a curve of growth analysis (Spitzer 1978) we estimated the 
hydrogen column densities implied by the observed equivalent 
widths of the two absorption lines at 8.05 and 9.79~keV.
Assuming the ion species responsible for the X-ray absorption in both lines is
Fe~{\sc xxv} and $b$ parameters of the order of the observed widths of the lines
($b = \sqrt{2}{\sigma_{u}}$, where $\sigma_{u}$ is the velocity width of the line), 
we calculate that the column densities of the absorbers
are $N_{Fe~{\sc XXV}abs1}$ $\approx$ (3.4$_{-0.7}^{+1.9}$) $\times$ 10$^{18}$~cm$^{-2}$ and
$N_{Fe~{\sc XXV}abs2}$ $\approx$ (3.8$_{-1.7}^{+2.9}$) $\times$ 10$^{18}$~cm$^{-2}$, respectively.
For no ionization correction and assuming solar abundances, the
implied total hydrogen column densities of these
absorbers are $N_{Habs1}$ $\approx$ (1.0$_{-0.2}^{+0.6}$) $\times$ 10$^{23}$~cm$^{-2}$ and
$N_{Habs2}$ $\approx$ (1.1$_{-0.5}^{+0.9}$) $\times$ 10$^{23}$~cm$^{-2}$, respectively.
We emphasize that there are significant limitations with
the present curve of growth analysis;
the absorption lines may contain multiple unresolved components 
implying that the equivalent widths and $b$ parameters used 
should be considered as upper limits. 
In addition, the velocity widths estimated from fits of Gaussian lines
to observed absorption features can only be used to
derive $b$ parameters when the absorber is optically
thin.
%The column densities that we find for the relativistic X-ray BALs
%are consistent within the errors with the total hydrogen
%column density (see Table 2) responsible for the observed 
%absorption (see left panel of Figure 1) at energies below 5~keV (rest frame).
It is not clear from the present data if the intrinsic
absorber which attenuates the low-energy continuum
and the absorber responsible for the absorption lines are the same.
The models adopted in Table 2 assume neutral absorption 
at $z = 3.91$. For an ionized absorber or for the case where 
there is partial covering, one expects the estimated hydrogen column density 
to be even larger than the value estimated assuming a neutral absorber
(i.e., a fit to the spectrum of \apm\ with an ionized absorber having an ionization parameter
of $U = 1$ yields a column density of $N_{\rm H}$ = 7.5 $\times$ 10$^{22}$~cm$^{-2}$).    
We note that the low-energy absorption could include
overlapping BALs from H-like and He-like ions of Mg, Si, and S
that perhaps could still survive along with the highly
ionized H-like and He-like ions of Fe (eg., Kallman \& Bautista 2001).
Based on our estimated column densities of the X-ray BALs, we calculated
the strengths of the associated absorption edges of Fe~{\sc xxv}
(Fe~{\sc xxvi}) to be $\tau_{Fe~{\sc XXV}}$ $\sim$ 0.07($\tau_{Fe~{\sc XXVI}}$ $\sim$ 0.04). 
We added these absorption edges to our spectral model for \apm\ (fit~3 of Table 2)
and found that we do not expect them to be detectable, consistent
with the data.

\begin{sloppypar}
The observed relativistic velocities, crude estimates of the
column densities, and locations of the X-ray BALs allow us to place constraints
on the mass-outflow rate for \apm. 
We estimated the mass-outflow rate as a function of the distance $R$ to the absorber
for $R/{\Delta}R$ ranging from 1--10, where ${\Delta}R$ is the thickness of the
absorber. We assumed a hydrogen column density of $N_{\rm H}$ = 1 $\times$ 10$^{23}$~cm$^{-2}$,
a covering fraction of 0.2, and a wind velocity of $0.2c$.
To obtain a reasonable mass-outflow rate (e.g., Proga et al. 2000) of $ < $ 1 M$_{\odot}$ year$^{-1}$
the radial location of the gas creating the X-ray BALs must be less than $\sim$ 10$^{17}$~cm
which is consistent with the radius implied from the
observed high velocities and ionization states of the X-ray BALs.  
\end{sloppypar}

To summarize, we have reported on the discovery of the first X-ray 
BALs in the gravitationally lensed quasar \apm.
The energies of the two observed absorption features 
of 8.05~keV and 9.79~keV (rest frame) suggest the presence of two 
distinct absorption systems with velocities of $\sim$ $0.2c$ and $\sim$ $0.4c$,
respectively. 
The combination of estimated short recombination time-scales for Fe~{\sc xxv} and/or Fe~{\sc xxvi}
and the observed relativistic velocities suggest that the X-ray absorbers are launched
at radii of $ < $ 2 $\times$ 10$^{17}$~cm.
Curve of growth estimates imply column
densities of the two X-ray BALS in \apm\ of the order of 1 $\times$ 10$^{23}$~cm$^{-2}$.
One of the key implications of our results is that X-ray BALs appear to be 
located within the UV BAL region and may therefore represent
the shielding gas proposed in several theoretical studies 
of line-driven disk winds (e.g., Murray et. al. 1995; Proga et al. 2000).  
We expect future follow-up observations of \apm\
with the \chandra\ high-energy transmission grating  and 
the \xmm\ RGS to identify the absorbing ions and resolve the X-ray absorption features into multiple components
allowing for tighter constraints of the properties of X-ray BALs. 
We also plausibly expect variability of the absorption-line 
profiles over short time-scales of $\sim$ days
based on our estimates of the launching radii of the X-ray absorbers.

During the review of this paper, a related paper was posted
on the ApJL web site by Hasinger, Schartel, \& Komossa (2002). They report 
the detection of an ionized Fe K edge in \apm\ with \xmm. 
A visual comparison between the \xmm\ and \chandra\ spectra of \apm\
suggests the presence of the two absorption features in both cases; however, the 
equivalent widths of the absorption features appear to be reduced 
during the deep \xmm\ observation. The edge model adopted in the analysis 
of the \xmm\ spectrum of \apm\ does not provide an acceptable fit to the \chandra\
spectrum of this object. This may imply significant variability of the 
X-ray BALs as suggested in our paper.\\

We thank Michael Eracleous and Doron Chelouche for useful discussions
and suggestions. We acknowledge financial support from NASA grants NAS 8-38252
and NAS 8-01128. WNB acknowledges financial support from NASA LTSA grant NAG5-8107 and
NASA grant NAG5-9932. SCG acknowledges financial support from NASA GSRP grant NGT5-50277
and from the Pennsylvania Space Grant Consortium.

\clearpage

\normalsize

\beginrefer

\refer Arav, N., Li, Z., \& Begelman, M.~C.\ 1994, \apj, 432, 62 \\

\refer Arav, N., Korista, K.~T., Barlow, T.~A., \& Begelman, M.~C.\ 1995, \nat, 
376, 576 \\

\refer Arnaud, K.~A.\ 1996, ASP 
Conf.~Ser.~101: Astronomical Data Analysis Software and Systems V, 5, 17 \\

\refer Allen, C. W., 1973, Astrophysical Quantities. Athlone Press, London \\

\refer Brandt, W.~N.~\& Schulz, N.~S.\ 2000, \apjl, 544, L123 \\

\refer Brinkmann, W., Ferrero, E., \& Gliozzi, M.\ 2002, \aap, 385, L31 \\

\refer Chartas, G., Dai, X., 
Gallagher, S.~C., Garmire, G.~P., Bautz, M.~W., Schechter, P.~L., \& 
Morgan, N.~D.\ 2001, \apj, 558, 119 \\

\refer Chelouche, D.~\& Netzer, H.\ 2001, \mnras, 326, 916 \\

\refer Ellison, S.~L., Lewis,
G.~F., Pettini, M., Sargent, W.~L.~W., Chaffee, F.~H., Foltz, C.~B.,
Rauch, M., \& Irwin, M.~J.\ 1999, \pasp, 111, 946 \\

\refer Egami, E., Neugebauer, 
G., Soifer, B.~T., Matthews, K., Ressler, M., Becklin, E.~E., Murphy, 
T.~W., \& Dale, D.~A.\ 2000, \apj, 535, 561 \\

\refer Fabian, A.~C.\ 1999, \mnras, 308, L39  \\

\refer Gallagher, S.~C., Brandt, W.~N., Laor, A., Elvis, M., Mathur, S., Wills, B.~J., \& Iyomoto, 
N.\ 2001, \apj, 546, 795 \\

\refer Gallagher, S.~C., Brandt, W.~N., Chartas, G., \& Garmire, G.~P.\ 2002, \apj, 567, 37 \\

\refer George, I.~M., Turner, 
T.~J., Yaqoob, T., Netzer, H., Laor, A., Mushotzky, R.~F., Nandra, K., \& 
Takahashi, T.\ 2000, \apj, 531, 52 \\

\refer Green, P.~J., Aldcroft, T.~L., Mathur, S., Wilkes, B.~J., \& Elvis, M.\ 2001, \apj, 558, 109 \\

\refer Hasinger, G., Schartel, N., \& Komossa, S.\ 2002, \apjl, 573, L77 \\  

\refer Ibata, R.~A., Lewis, 
G.~F., Irwin, M.~J., Leh{\' a}r, J., \& Totten, E.~J.\ 1999, \aj, 118, 1922 \\

\refer Kaspi, S., Brandt, W. N., George, I. M., Netzer, H., Crenshaw, D. M., 
Gabel, J. R., Hamann, F. W., Kaiser, M. E., Koratkar, A., Kraemer, S. B., Kriss, G. A.,
Mathur, S., Mushotzky, R. F., Nandra, K., Peterson, B., 
Shields, J. C., Turner, T. J., \& Zheng W.\ 2002, astro-ph/0203263 \\

\refer Kallman, T.~\& Bautista, M.\ 2001, \apjs, 133, 221 \\

\refer Kotani, T., Ebisawa, K., 
Dotani, T., Inoue, H., Nagase, F., Tanaka, Y., \& Ueda, Y.\ 2000, \apj, 
539, 413 \\

\refer Laor, A.~\& Brandt, 
W.~N.\ 2002, \apj, 569, 641 \\

\refer Lucy, L.~B.\ 1974, \aj, 79, 745 \\

\refer Mu{\~n}oz, J.~A., Kochanek, C.~S., \& Keeton, C.~R.\ 2001, \apj, 558, 657 \\

\refer Murray, N., Chiang, J., Grossman, S.~A., \& Voit, G.~M.\ 1995, \apj, 451, 498 \\

\refer Nandra, K., George, 
I.~M., Mushotzky, R.~F., Turner, T.~J., \& Yaqoob, T.\ 1999, \apjl, 523, L17 \\

\refer Oshima, T., Mitsuda, K., Fujimoto, R., Iyomoto, N., 
Futamoto, K., Hattori, M., Ota, N., Mori, K.,  
Ikebe, Y., Miralles, J.~M., \& Kneib, J.-P., \ 2001, \apjl, 563, L103 \\

\refer Proga, D., Stone, J.~M., \& Kallman, T.~R.\ 2000, \apj, 543, 686 \\

\refer Reeves, J.~N.~\& Turner, M.~J.~L.\ 2000, \mnras, 316, 234 \\

\refer Richardson, W.~H.\ 1972, Optical Society of America Journal, 62, 55 \\

\refer Spitzer, L. 1978, Physical Processes in the Interstellar Medium (New York: Wiley)\\

\refer Srianand, R.~\& Petitjean, P.\ 2000, \aap, 357, 414 \\

\refer Srianand, R., Petitjean, P., Ledoux, C., \& Hazard, C.\ 2002, 
MNRAS, in press, astro-ph/0205524 \\

\refer Stark, A.~A., Gammie, 
C.~F., Wilson, R.~W., Bally, J., Linke, R.~A., Heiles, C., \& Hurwitz, M.\ 
1992, \apjs, 79, 77 \\

\refer Turnshek, D.~A., Grillmair, C.~J., Foltz, C.~B., \& Weymann, R.~J.\ 1988, 
\apj, 325, 651 \\

\refer Tsunemi, H., Mori, K., 
Miyata, E., Baluta, C., Burrows, D.~N., Garmire, G.~P., \& Chartas, G.\ 
2001, \apj, 554, 496 \\

\refer Ueda, Y., Asai, K., 
Yamaoka, K., Dotani, T., \& Inoue, H.\ 2001, \apjl, 556, L87 \\

\refer Weymann, R.~J., Morris, S.~L., Foltz, C.~B., \& Hewett, P.~C.\ 1991, \apj, 
373, 23 \\

\refer Wise, M. W., Davis, J. E., Huenemoerder, Houck, J. C., Dewey, D.
Flanagan, K. A., and Baluta, C. 1997,
{\it The MARX 3.0 User Guide, CXC Internal Document}
available at http://space.mit.edu/ASC/MARX/ \\

\endrefer

\clearpage
\scriptsize
\begin{center}
\begin{tabular}{clclllll}
\multicolumn{8}{c}{TABLE 1} \\
\multicolumn{8}{c}{Log of Observations of \apm} \\
& & & & & & &  \\ \hline\hline
\multicolumn{1}{c} {Observation} &
\multicolumn{1}{c} {Obsid} &
\multicolumn{1}{c} {Exposure} &
\multicolumn{1}{c} {$R_{\rm A}$$^{a}$} &
\multicolumn{1}{c} {$R_{\rm B}$$^{a}$} &
\multicolumn{1}{c} {$R_{\rm C}$$^{a}$} &
\multicolumn{1}{c} {$R_{\rm tot}$$^{b}$} &
\multicolumn{1}{c} {$R_{\rm Bkg}$$^{c}$} \\
\multicolumn{1}{c} {Date} &
\multicolumn{1}{c} {} &
\multicolumn{1}{c} {Time} &
\multicolumn{1}{c} {} &
\multicolumn{1}{c} {} &
\multicolumn{1}{c} {} &
\multicolumn{1}{c} {} &
\multicolumn{1}{c} {} \\
              &     & (s) & 10$^{-2}$ cnts s$^{-1}$ & 10$^{-2}$ cnts s$^{-1}$ & 10$^{-2}$ cnts s$^{-1}$ & 10$^{-2}$ cnts s$^{-1}$&10$^{-6}$ cnts s$^{-1}$ pixel$^{-1}$ \\ \hline
                  &     &      &         &   &     &  &\\
2000 Oct 11 & 1643 &  9,137 & {$\ldots\ldots\ldots$}$^{d}$  &  {$\ldots\ldots\ldots$}$^{d}$   &     {$\ldots\ldots\ldots$}$^{d}$                                        & 10.96 $\pm$ 0.34 & 1.68 $\pm$ 0.22\\
2002 Feb 24 & 2979 & 88,817 & 3.1 $\pm$ 0.07   & 2.8 $\pm$ 0.06   & 0.66 $\pm$ 0.07 & 6.56 $\pm$ 0.09 &  0.99 $\pm$ 0.05\\
                  &     &      &         &    &    & &\\
\hline
\end{tabular}
\end{center}
NOTES:\\
\noindent
{}$^{a}$ $R_{\rm A}$, $R_{\rm B}$ and $R_{\rm C}$  are the event rates for images A, B and C estimated 
from a two-dimensional fit to the \chandra\ image of \apm.
Only events with standard \asca\ grades 0,2,3,4,6
and with energies from 0.2--10~keV were used in the
binned image. \\

\noindent
{}$^{b}$ $R_{\rm tot}$ is the detected event rate from \apm\
extracted from a circular region centered on the midpoint of images A and B
with a radius of 3{\arcsec}. Only events with standard \asca\ grades 0,2,3,4,6
and with energies from 0.2--10~keV were extracted. \\

\noindent
{}$^{c}$ $R_{\rm Bkg}$ is the detected background event rate per ACIS pixel (1 ACIS pixel = 0{\sarc}492) extracted
from an annulus centered on \apm\ with inner and outer radii of 10{\arcsec} and 30{\arcsec}, respectively.
Only events with standard \asca\ grades 0,2,3,4,6
and with energies from 0.2--10~keV were extracted. \\

\noindent
{}$^{d}$ It was not possible to determine the individual count rates 
for the first observation of \apm\ due to the 
low S/N and the presence of moderate photon pile-up. \\

\clearpage
\scriptsize
\begin{center}
\begin{tabular}{clcc}
\multicolumn{4}{c}{TABLE 2}\\
\multicolumn{4}{c}{RESULTS FROM FITS TO THE COMBINED} \\
\multicolumn{4}{c}{SPECTRUM OF ALL IMAGES OF \apm} \\
 & & &  \\ \hline\hline
\multicolumn{1}{c} {Fit$^{a}$} &
\multicolumn{1}{c} {Model} &
\multicolumn{1}{c} {Parameter$^{b}$} &
\multicolumn{1}{c} {Value$^{c}$} \\
  &                               &                   &                                                  \\ \hline
  &                               &                   &                                                   \\
1 &  Power-Law (PL) and           &   $\Gamma_{AB}$   &1.72$_{-0.06}^{+0.03}$                            \\
  &  neutral absorption at        &   $N_{\rm H,AB}$  & (6.0$_{-0.8}^{+0.8}$) $\times$ 10$^{22}$~cm$^{-2}$ \\
  &  source.                      &   $\chi^2/{\nu}$  & 182.1/109    \\
  &                               &   $P(\chi^2/{\nu})$$^{d}$  &    1.4 $\times$ 10$^{-5}$                                      \\
  &                               &                   &                                                   \\
  &                               &                   &                                                   \\
2 &  PL, neutral absorption       & $\Gamma_{AB}$     &1.73$_{-0.06}^{+0.06}$                            \\
  &  at source, and one           & $N_{\rm H,AB}$    &(6.4$_{-0.9}^{+0.8}$) $\times$ 10$^{22}$~cm$^{-2}$ \\
  &  Gaussian absorption line     &  E$_{abs1}$       & 8.05$_{-0.07}^{+0.18}$~keV    \\
  &  at source.                   &  $\sigma_{abs1}$  &$ < $ 0.140~keV                                       \\
  &                               &  EW$_{abs1}$      & 0.23$_{-0.07}^{+0.06}$~keV                         \\
  &                               &  $\chi^2/{\nu}$   &    146.9/106                           \\
  &                               &  $P(\chi^2/{\nu})$$^{d}$ &   5.3 $\times$ 10$^{-5}$            \\
  &                               &                   &                                                  \\
3 &  PL, neutral absorption       & $\Gamma_{AB}$     & 1.72$_{-0.05}^{+0.06}$                           \\
  &  at source, and two           &  $N_{\rm H,AB}$   & (6.7$_{-0.8}^{+0.9}$)$\times$ 10$^{22}$~cm$^{-2}$   \\
  &  Gaussian absorption lines    &  E$_{abs1}$       & 8.05$_{-0.08}^{+0.10}$~keV     \\
  &  at source.                   & $\sigma_{abs1}$   & $ < $ 0.140~keV     \\
  &                               &  EW$_{abs1}$      & 0.24$_{-0.07}^{+0.06}$~keV                      \\
  &                               &  E$_{abs2}$       & 9.79$_{-0.19}^{+0.20}$~keV              \\
  &                               &  $\sigma_{abs2}$  & 0.41$_{-0.16}^{+0.19}$~keV                \\
  &                               &  EW$_{abs2}$      & 0.43$_{-0.13}^{+0.17}$~keV                    \\
  &                               &  $\chi^2/{\nu}$   & 106.7/103                                \\
  &                               &$P(\chi^2/{\nu})$$^{d}$ & 0.41                                \\
  &                               &                   &                                          \\
\hline \hline
\end{tabular}
\end{center}

NOTES-\\
${}^{a}$ All model fits include fixed, Galactic absorption of
$N_{\rm H}=3.9\times10^{20}$~cm$^{-2}$ (Stark et al. 1992). \\
${}^{b}$ All absorption-line parameters are calculated for the rest frame.\\ 
${}^{c}$ All errors are for 90\% confidence with all
parameters taken to be of interest except absolute normalization.\\
${}^{d}$ $P(\chi^{2}/{\nu})$ is the probability of exceeding $\chi^{2}$
for ${\nu}$ degrees of freedom.

\clearpage
\begin{figure*}
\centerline{\includegraphics[width=18cm]{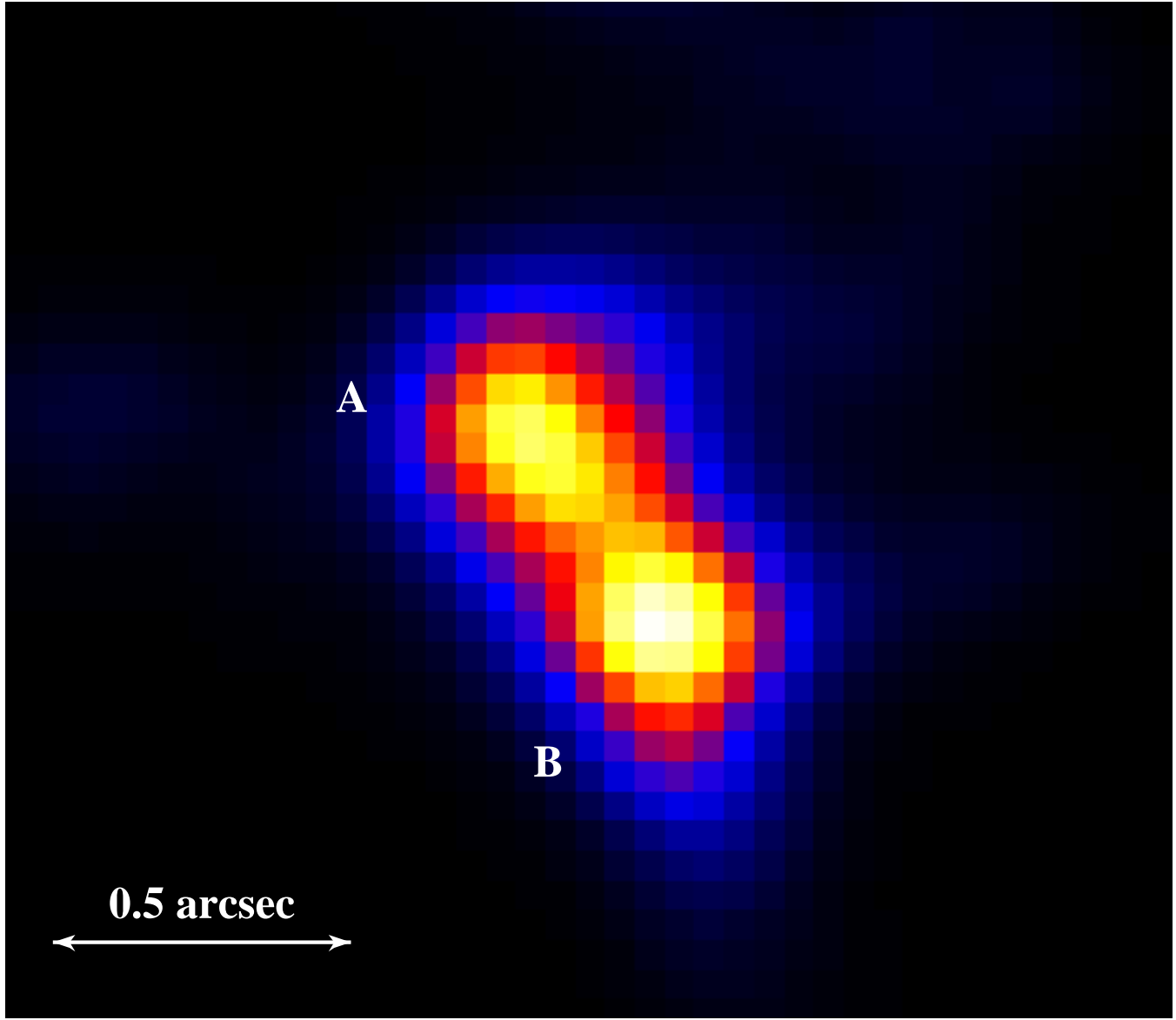}}
\caption{ \small Deconvolved image of the 2002 Feb 24 \chandra\ observation of \apm.
North is up, and east is to the left.
\label{f1.ps}}
\end{figure*}

\clearpage
\begin{figure*}
\centerline{\includegraphics[width=18cm]{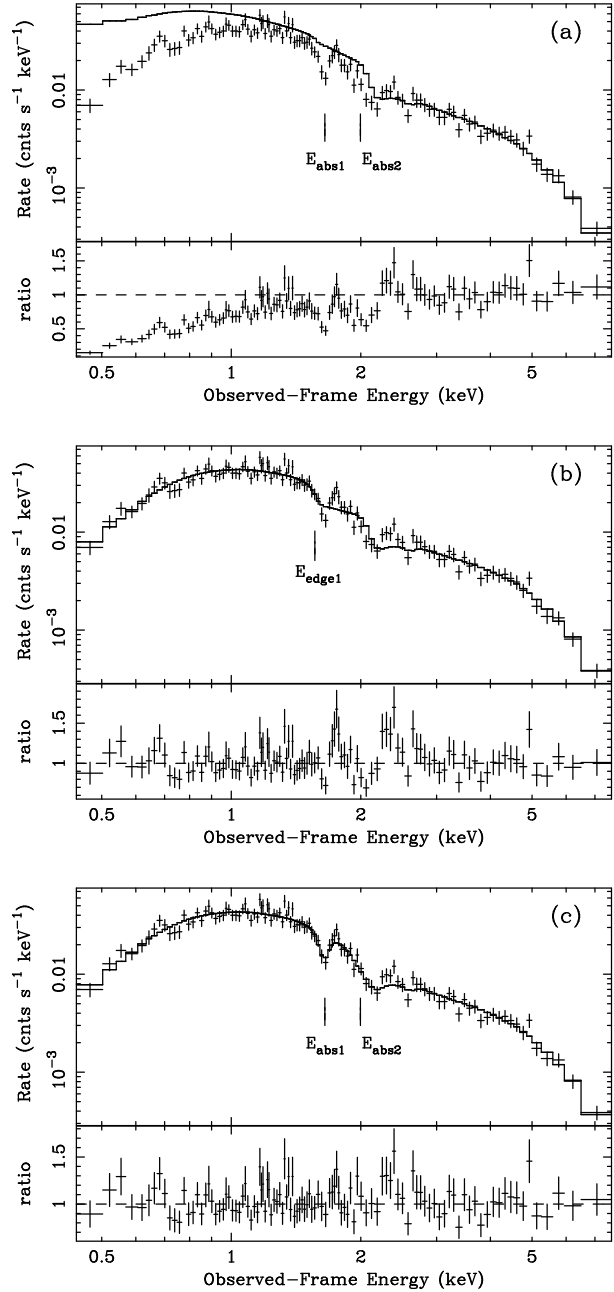}}
\caption{ \small (a) The top panel shows the \chandra\ observed-frame spectrum of
the combined images of \apm\ fit with Galactic absorption and a
power-law model above 2.2~keV (10.8~keV rest frame)
that is extrapolated to lower energies.
The lower panel represents the ratio of the data to the model.
Two absorption features within 1.5--3.0~keV
are noticeable in the ratio plot.
(b) The top panel shows the \chandra\ spectrum of
the combined images of \apm\ fit with Galactic absorption,
neutral absorption at the source, a power-law model,
and an absorption edge. In the lower panel the ratio plot of data to model
indicates the presence of significant residuals.
This model did not provide an acceptable fit.
(c) The top panel shows the \chandra\ spectrum of
the combined images of \apm\ fit with Galactic absorption,
neutral absorption at the source,
a power-law model, and two Gaussian absorption lines.
In the lower panel the ratio plot of data to model for fit 3 of Table 2
indicates that this model can account for
most of the spectral features in \apm.
\label{f2.ps}}
\end{figure*}

\clearpage
\begin{figure*}
\centerline{\includegraphics[width=18cm]{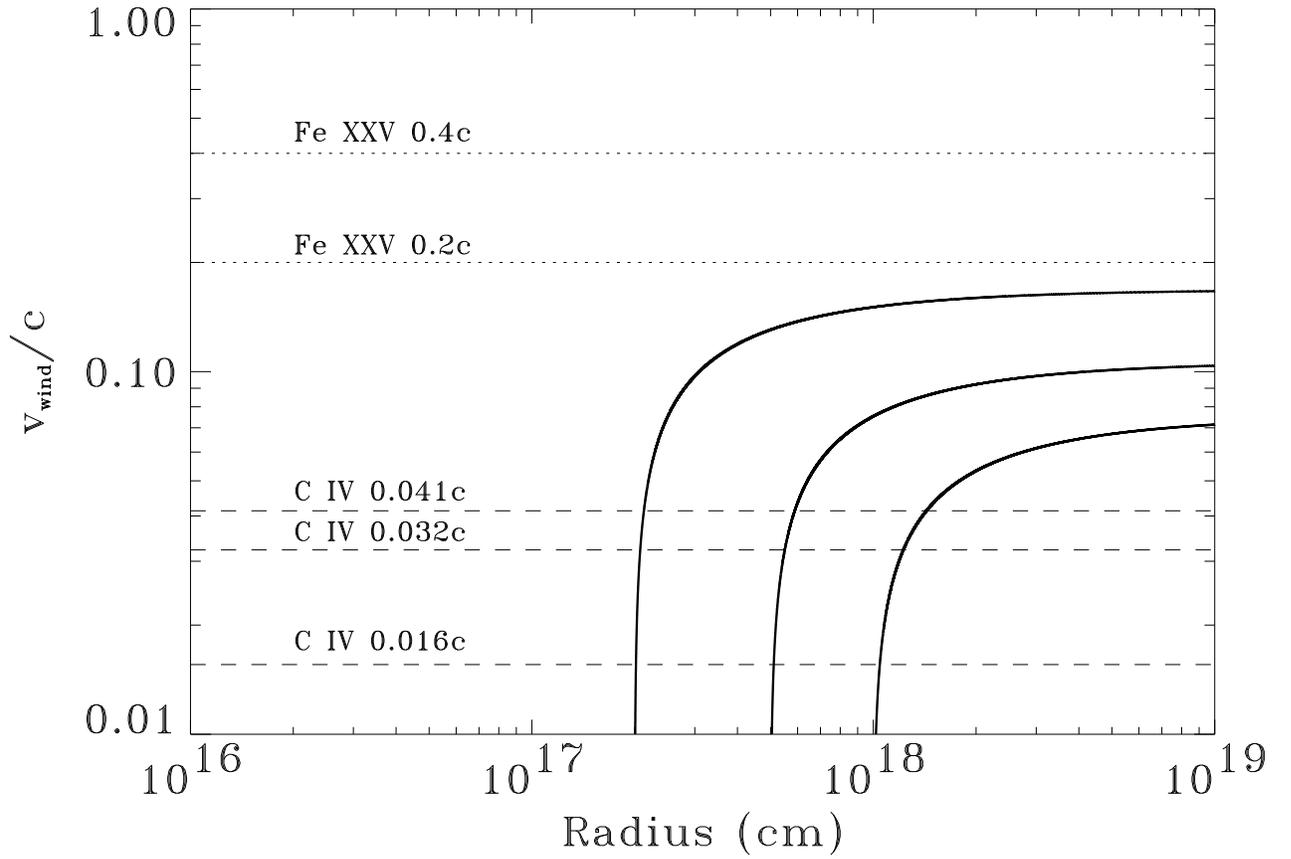}}
\caption{ \small Wind velocity as a function
of radius from the central source for a radiation pressure driven wind.
For a qualitative comparison we have estimated
the wind velocities  for launching
radii of 2 $\times$ 10$^{17}$~cm, 5 $\times$ 10$^{17}$~cm,
and 1 $\times$ 10$^{18}$~cm. We have over-plotted
the observed C~{\sc iv} BAL (dashed lines) and Fe~{\sc xxv} BAL (dotted lines) velocities.
\label{f3.ps}}
\end{figure*}

\end{document}